\documentclass[twocolumn,showpacs,preprintnumbers]{revtex4}
\usepackage{amsmath}
\usepackage{amssymb}
\usepackage{graphicx}
\usepackage{bm}
\begin{document}
\title{Break-up mechanisms in heavy ion collisions at low energies}
\author{L. Shvedov$^{a}$, M. Colonna$^{a}$, M. Di Toro$^{a,b}$} 
\affiliation{
        $^a$LNS-INFN, I-95123, Catania, Italy\\
	$^b$Physics and Astronomy Dept. University of Catania, Italy\\}

\begin{abstract}
We investigate reaction mechanisms occurring in heavy ion collisions
at low energy (around 20 MeV/u). In particular, we focus on the competition
between fusion and break-up processes (Deep-Inelastic and fragmentation) 
in semi-peripheral collisions, where the formation of excited systems in various 
conditions of shape and angular momentum is observed. 
Adopting a Langevin treatment for the dynamical evolution of the system
configuration, described in terms of shape observables 
such as quadrupole and octupole moments, we derive fusion/fission probabilities,
from which one can finally evaluate the corresponding fusion and break-up cross sections. 
The dependence of the results on shape, angular momentum and excitation energy 
is discussed.  


\end{abstract}
\maketitle

\section{Introduction}

Nuclear reactions between medium-mass nuclei at low energies (around 20 MeV/u)
offer the possibility to investigate several aspects of dissipative mean-field
dynamics and to probe nuclear matter under extreme conditions with
respect to shape, spin and excitation energy. 
In this energy domain, well above the Coulomb barrier but below the Fermi energies, 
one essentially observes two types of reaction mechanisms:
Fusion dominates in the case of central and semi-peripheral collisions, while binary reseparation processes, 
associated with deep-enelastic or fast fission mechanisms, essentially involve the remaining 
range of (semi-)peripheral reactions \cite{Laut06}. 
However, along the transition from fusion to binary processes, composite systems
with rather elongated shape and large intrinsic angular momentum can be formed, corresponding to
metastable (or even unstable) conditions, where mean-field fluctuations may play a decisive role 
in determining the final outcome. 
The presence of large event by event variances related to the onset of new instabilities have been
already noted in experiments, from the anomalous distribution of primary fragment properties
in binary events \cite{Aw84,Pen90}. The observed variances (in mass, charge, excitation energy,
angular distribution) appeared much larger than the ones predicted by mean-field nucleon exchange
models. Similar conclusions were reached in theory simulations based on stochastic transport
models \cite{Col95}.

Interaction times are quite long and a large coupling among various mean-field modes is expected, 
leading to a co-existence of the different reaction mechanisms in semi-central collisions.   
The study of the competition between fusionlike and binarylike processes and, more
generally, of the fate of the hot nuclear residues created in these reactions
is a longstanding problem, from which one can learn
a lot about mean-field dynamics and fundamental properties of nuclear forces. 
This issue has recently
found a renewed interest, due to the possibility to perform new analyses involving neutron-rich
or even exotic systems \cite{Amo09}. In these conditions the reaction mechanism characterizing dissipative
collisions is expected to be sensitive to the density dependence of the isovector part of the
nuclear interaction, a matter that is largely debated nowadays \cite{Amo09,rep}.  

In reactions involving medium-heavy nuclei, as a result of the complex neck dynamics, one can
also observe, in sufficiently inelastic collisions, new modes of reseparation of the colliding
system, such as dynamical ternary breaking, 
with massive fragments nearly aligned along a common separation
axis \cite{Gla,Stef}. Experimental evidences of this mechanism
have been recently reported in the case of $^{197}$Au + $^{197}$Au collisions at 15 MeV/u,
where also aligned quaternary breaking has been observed \cite{Isa}.     
These effects could still be explained in terms of the persistence of the excitation of shape
and rotational modes in the projectile-like(PLF) and/or target like(TLF) fragments that are formed
in binarylike events, that would lead to further reseparation along a preferential axis,
similarly to what happens in fast-fission processes of PLF or TLF. 
It is worth mentioning that, at higher beam energy (around 40 MeV/u), 
where apart from mean-field effects 
two-body correlations are important,
ternary breakings become the dominant process and
new features are observed, corresponding to the emission
of small fragments coming directly from the strongly interacting neck region \cite{Def}.
Actually one may think in terms of a smooth transition between the different decay modes of
PLF and/or TLF, from fast fission, characterized by the splitting into fragments with
similar size and small relative velocity, to neck emission, where small fragments are emitted 
with larger relative velocity with respect to PLF and TLF.

From the above discussion, it is clear that the understanding of the competition between reaction
mechanisms in dissipative collisions, as well as of the nature of new exotic reseparation modes, 
requires a thorough analysis of the underlying mean-field dynamics and associated shape
fluctuations and rotational effects. 
In this paper, we attempt to improve the dynamical description of low energy collisions by
coupling a microscopic transport approach based on mean-field concepts, suitable to follow
the early stage of the collision up to the formation of composite excited sources, 
to  a more refined treatment  of the dynamics of 
shape observables, including the associated fluctuations within
the Langevin scheme \cite{David}, for the following evolution up to the definition of the final outcome. 
In particular, we will discuss the dynamics of excited  sources characterized by
given values of quadrupole and octupole moments and intrinsic angular momentum. 
This allows one to investigate  
the competition between
fusionlike and binarylike reaction mechanisms and
to evaluate fusion cross sections,  as well as the probability and the features  
of fast-fission processes of PLF (or TLF).
The paper is organized as it follows: In Section 2 we present the hybrid transport treatment
employed to 
follow the dynamical evolution of the system. 
Results concerning the competition between
fusion and binary processes are discussed in Section 3.  Finally conclusions and perspectives
are drawn in Section 4.

\section{Simulation of the collisional dynamics}

\subsection{Dynamical description of nuclear reactions}
The evolution of systems governed  by a complex phase space
can be described
by  a transport equation, of the Boltzmann-Nordheim-Vlasov (BNV) type, with a fluctuating term, the so-called
Boltzmann-Langevin equation (BLE) \cite{Ayik,BL_new}:
\begin{equation}
{{df}\over{dt}} = {{\partial f}\over{\partial t}} + \{f,H\} = I_{coll}[f] 
+ \delta I[f],
\label{BL}
\end{equation}
 where $f({\bf r},{\bf p},t)$ is the one-body distribution function, 
or Wigner transform of the one-body density, 
$H({\bf r},{\bf p},t)$ the mean field Hamiltonian, 
$I_{coll}$ the two-body collision term (that accounts for the residual interaction) 
incorporating the Fermi statistics of the particles,
and 
$\delta I[f]$ the fluctuating part of the
collision integral.
The nuclear EOS, directly linked to the mean-field Hamiltonian $H$, can be written as: 
\begin{equation}
E/A(\rho,I) = E_s/A(\rho) +  C_{sym}(\rho)I^2 + O(I^4),
\end{equation}
where $I=(N-Z)/A$ is the asymmetry parameter.
We adopt a soft isoscalar EOS, $E_s/A(\rho)$, with compressibility modulus $K = 200~MeV$, 
which is favored e.g. from flow studies \cite{DanLac}.
For the density  ($\rho$) dependence of the symmetry energy, $C_{sym}(\rho)$,
we consider a linear increase of the potential  part
of the symmetry energy with density (asystiff):
\begin{equation}
C_{sym}\left( \rho \right) =a\cdot \left( \frac{\rho }{\rho _{0} } \right) ^
{2/3}
+ b \cdot (\rho / \rho_0),
\label{eq3new}
\end{equation}
where $\rho_0$ is the saturation density, a=13.4 MeV and b=18 MeV.
From the expression of the energy density, Eq.(2), the mean-field potential is directly derived. 
The free energy- and angle-dependent nucleon-nucleon cross section is used
in the collision integral \cite{TWINGO}. 

Within such approach, 
the system is described in terms of the one-body distribution function $f$, but this function
may experience a stochastic evolution in response to the action of the fluctuating term $\delta I[f]$.

However, 
the numerical resolution of the full BL equation is not available yet in 3D. 
Approximate treatments to the BLE have been introduced so far, see 
Refs.\cite{TWINGO,Salvo}, such as  
the Stochastic Mean Field (SMF) model, 
that consists in the implementation of stochastic density fluctuations only in coordinate space
and can be solved numerically
using the test particle method \cite{TWINGO}. 
The latter  approach has shown to be particularly appropriate for the 
description of the evolution of the dilute unstable sources that develop
in dissipative collisions at Fermi energies (30-100 MeV/u) \cite{rep1}. 
However, here we are essentially interested in semi-central reactions at lower energies
where, most likely, the formation of elongated (rather than dilute) systems 
is observed, and phenomena associated with surface
(rather than volume) metastabity and/or instability may take place. 
To improve the treatment of fluctuations suitable to describe the latter scenario, we will adopt
a hybrid description of the dynamics: 
We follow the miscroscopic SMF evolution until the time instant when 
local thermal equilibrium is established and one observes the formation of
quasi-stationary elongated systems, with density close to the normal value.
Then, to deal with the following evolution of the system, we move to a more macroscopic 
model description, where the system is characterized in terms of global observables, 
for which the full treatment of fluctuations in phase space 
is numerically affordable, as explained below.

\subsection{Dynamical evolution of shape observables}
This Section is devoted to the description of the dynamical evolution of 
excited systems whose leading degrees of freedom are shape observables, 
while the density keeps always close to the normal value, $\rho_0 = {{3}\over{4\pi r_0^3}}$,
being $r_0$ the nuclear radius constant ($r_0 = 1.2~fm$). 
The configuration of the system under study, having given charge Z and mass A,  
is described by three global observables (and associated velocities):
the quadrupole moment $\beta_2$, the octupole moment $\beta_3$ 
and the rotation angle $\omega$.
For situations far from the spherical shape
the thermal agitation can induce fluctuations that may eventually lead to
break-up channels. Hence the correct treatment of shape fluctuations 
is crucial for the characterization of the reaction mechanism.
To this purpose,  we consider the stochastic extension of 
the Rayleigh-Lagrange equations of motion \cite{Bl3} 
(the Langevin equation):
\begin{equation}
\frac{\mathrm{d}}{\mathrm{d}t}\frac{\partial L}{\partial \dot{q}_i}+\frac
{\partial F}{\partial \dot{q}_i}=\frac{\partial L}{\partial q_i}+F_{fluc}(t),
\label{RLE}
\end{equation}
where $q_i (i = 1,2,3) = (\omega,\beta_2,\beta_3)$.
$L(q_i,\dot{q}_i)=E_{kin}(q_i,\dot{q}_i)+E_{rot}(q_i,\dot{q}_i)-E_{pot}
(q_i)$ denotes the Lagrangian of the system and
\begin{equation}
F(q_i,\dot{q}_i)=\frac{1}{2}\frac{\mathrm{d}E_{tot}}{\mathrm{d}t}=\frac{1}{2}\sum_
{i,j=2}^3R_{ij}\dot{q}_i\dot{q}_j
\end{equation}
is the Rayleigh dissipation function. 
$E_{kin}$, $E_{rot}$ and $E_{pot}$ indicate the kinetic, rotational and potential energy
of the system, respectively, and the  quantity $R_{ij}$ is the
dissipation tensor. 
The difference with respect to 
the standard Rayleigh-Lagrange equations is the fluctuation term $F_{fluc}$, 
that can be interpreted as a rapidly
fluctuating stochastic force, in the same spirit of the Brownian motion,
similar to the fluctuating term of the BLE, Eq.(1).   
We solve numerically the set (\ref{RLE}) of coupled equations. 

For given values of the quadrupole and octupole moments, 
the shape of the system is parametrized, in terms of the polar angle $\theta$, as it follows:
$$
R(\theta)=R_0(\beta_2,\beta_3)\{1+\beta_1(\beta_2,\beta_3)Y_{10}
(\theta)+
$$
\begin{equation}
+\beta_2Y_{20}(\theta)+\beta_3Y_{30}(\theta) \},
\end{equation}
where the functions $Y_{i0}(\theta)$ 
are spherical harmonics.
The parameters $\beta_1$  and $R_0$ are introduced to 
conserve the position of the center of mass and the total volume V of the system and
can be determined from the equations:
\begin{equation}
\int dV z = \frac{2\pi}{4} \int_0^\pi R^4(\theta)\sin\theta\cos\theta d\theta=0,
\end{equation}
\begin{equation}
\int dV=  \frac{2\pi}{3} \int_0^\pi R^3(\theta)\sin\theta d\theta = \frac{4}{3}\pi r_0^3A,
\end{equation}
where $z$ denotes the coordinate along the system maximum elongation axis (or symmetry axis). 
In the following we discuss in detail the derivation of the 
different terms of the Lagrangian $L$.  

\subsubsection{Rotational energy}
The rotational energy is simply equal to:
\begin{equation}
E_{rot}=\frac{1}{2}I(\beta_2,\beta_3)\dot{\omega}^2,
\end{equation}
where
\begin{equation}
I(\beta_2,\beta_3)=\frac{\pi m\rho_0}{5}\int^\pi_0 R^5(\theta)\{1+\cos^2\theta\}\sin\theta
\,\mathrm{d}\theta
\end{equation}
is the moment of inertia for the whole system,
being $m$ the nucleon mass.

\subsubsection{Kinetic energy}
The kinetic energy can be expressed as it follows:
\begin{equation}
E_{kin}=\frac{1}{2}\sum_{i,j=2}^3M_{ij}(\beta_2,\beta_3)\dot{q}_i\dot{q}_j
\end{equation}
To calculate the mass tensor $M_{ij}$, we adopt the prescriptions of Ref.\cite{Ni1}:
\begin{equation}
M_{i,j} = \frac{1}{2}(M_{i,j}'+M_{j,i}')
\end{equation}
with
\begin{eqnarray}
M_{i,j}'=2\pi m\rho_0\int^\pi_0\sum^L_{l=1}b_{il}R^{l+2}(\theta)P_l(\cos\theta)
\{ \frac{\partial R_0}{\partial\beta_j}S+
\nonumber\\
+R_0\left(\frac{\partial\beta_1}{\partial\beta_j}
Y_{10}+Y_{j0}\right) \}\sin\theta\,\mathrm{d}\theta.
\end{eqnarray}
Here $P_l$ are Legendre polynomials and $S(\theta)=\frac{R(\theta)}{R_0}$. In our calculations
we have $L=5$. 
The coefficients $b_{2l}$ and $b_{3l}$ are obtained solving the system of equations:
\begin{eqnarray}
\sum^L_{l=1}A_{kl}b_{ml}=C_{mk} & & k=1\dots L, ~~m = 2,3
\end{eqnarray}
with
$$
A_{kl}=\int^\pi_0R^{l-1}(\theta)\left\{lP_l(\cos\theta)-\frac{1}{R(\theta)}\frac{\partial R(\theta)}
{\partial\theta}\frac{\partial P_l(\cos\theta)}{\partial\theta}\right\}
$$
\begin{equation}
\cdot R^{k-1}(\theta)\left\{kP_k(\cos\theta)-\frac{1}{R(\theta)}\frac{\partial R(\theta)}
{\partial\theta}\frac{\partial P_k(\cos\theta)}{\partial\theta}\right\}\sin\theta\,\mathrm{d}\theta,
\end{equation}
$$
C_{mk}=\int^\pi_0R^{k-1}(\theta)\left\{kP_k(\cos\theta)-\frac{1}{R(\theta)}\frac{\partial R(\theta)}
{\partial\theta}\frac{\partial P_k(\cos\theta)}{\partial\theta}\right\}
$$
\begin{equation}
\cdot\left\{\frac{\partial R_0}{\partial\beta_m}S+R_0\left(\frac{\partial\beta_1}{\partial\beta_m}
Y_{10}+Y_{m0}\right)\right\}\sin\theta\,\mathrm{d}\theta.
\end{equation}

\subsubsection{Potential energy: Nuclear term}

\noindent
Concerning the nuclear part of the potential energy, $E_n$, 
we discuss essentially the surface contribution, since our system keeps a  
volume constant in time. 
We adopt a double volume integral of the
Yukawa-plus-exponential folding function \cite{Kr}:
\begin{equation}
E_n=-\frac{a_s(1-k_sI^2)}{8\pi^2r_0^{\phantom{0}2}a^3}\int_V\int_V\left(\frac
{\sigma}{a}-2\right)\frac{e^{-\sigma/a}}{\sigma}\,\mathrm{d}^3r\,\mathrm{d}^3r',
\label{pot_nuc}
\end{equation}
where 
$a_s$ is the surface-energy constant, $k_s$
is the surface-asymmetry constant and $a$
is the range of the Yukawa-plus-exponential potential. 
 $\sigma$ denotes the modulus of the relative distance $\sigma =|r-r'|$. 
Parameters have been fitted to the
ground-state energies and fission barrier heights \cite{Mo1,Mo2}. In order to reduce
the numerical efforts, 
the integral of Eq.(\ref{pot_nuc}) can be 
transformed into a double surface integral, by using the
twofold Gauss divergence theorem. For axially symmetric shapes, one of the azimuthal
integrations can be performed trivially \cite{Kr,Si1} and the resulting threefold
integral is:
$$
E_n=\frac{a_s(1-k_sI^2)}{4\pi r_0^{\phantom{0}2}}\iiint\left\{2-\left[\left(
\frac{\sigma}{a}\right)^2+2\frac{\sigma}{a}+2\right]e^{-\sigma/a}\right\}
$$
\begin{equation}
\times\frac{P(\theta,\theta',\phi)P(\theta',\theta,-\phi)}{\sigma^4}\mathrm{d}\theta\,
\mathrm{d}\theta'\,\mathrm{d}\phi,
\end{equation}
where the distance $\sigma$ can be expressed as:
$$
\sigma=[R^2(\theta)+R^2(\theta')-2R(\theta)R(\theta')
$$
$$
\cdot\{\cos\theta\cos\theta'+\sin\theta
\sin\theta'\cos\phi\}]^{1/2}
$$
and
$$
P(\theta,\theta',\phi)=R(\theta)\sin\theta\{R^2(\theta)-R(\theta)R(\theta')[\cos\theta
\cos\theta'+
$$
$$
+\sin\theta\sin\theta'\cos\phi]-R(\theta')\frac{\partial R(\theta)}{\partial\theta}[\sin\theta\cos\theta'
-\cos\theta\sin\theta'\cos\phi]\}
$$

\subsubsection{Potential energy: Coulomb term}

The Coulomb part of the potential energy is taken as \cite{Da2}:
\begin{displaymath}
E_C=E_C^{sharp}+\Delta E_C^{dif}
\end{displaymath}
where $E_C^{sharp}$ is the Coulomb energy corresponding to a sharp charge density distribution and
$\Delta E_C^{dif}$ is a correction due to the diffuseness. 

The sharp-surface part of the Coulomb energy is equal to:
\begin{equation}
E_C^{sharp}=-\frac{\rho_{p}^2\pi}{6}\iiint\frac{P(\theta,\theta',\phi)
P(\theta',\theta,-\phi)}{\sigma}\,\mathrm{d}\theta\,\mathrm{d}\theta'\,\mathrm{d}\phi
\end{equation}
where $\rho_{p}$ is the charge(proton) density, $\rho_p = {Z\over A} \rho_0$. 
The correction to the Coulomb energy due to the
diffuseness can be expressed as:
$$
\Delta E_C^{dif}=\rho_{p}^{\phantom{ch}2}\pi a_C^{\phantom{C}3}\iiint \{
2\frac{\sigma}{a_C}-5+\left[\frac{1}{2}\left(\frac{\sigma}{a_C}\right)^2+3
\frac{\sigma}{a_C}+5\right]
$$
\begin{equation}
\times e^{-\sigma/a_C} \}\frac{P(\theta,\theta',\phi)P(\theta',\theta,-\phi)}{\sigma^4}\,\mathrm{d}\theta\,
\mathrm{d}\theta'\,\mathrm{d}\phi,
\end{equation}
where $a_C$ is the range parameter of the Yukawa function generating the diffuse charge
distribution \cite{Da2,Si1,My1}. 

\subsubsection{Dissipation function}
The one-body dissipation mechanism is evaluated as it follows
 (see Ref.\cite{Bl3} for details): 
\begin{equation}
\frac{\mathrm{d}E}{\mathrm{d}t}=m\rho_0\overline{v}\oint\dot{n}^2\,\mathrm{d}S
\end{equation}
where the integration is performed over the whole surface of the system, 
$\overline{v}=\frac{3}{4}v_F$ is the average nucleon
velocity and
\begin{eqnarray}
\dot{n}^2=\frac{\left|\frac{\partial {\cal R}}{\partial t}\right|^2}{\left|\nabla {\cal R}\right|^2}, & & 
{\cal R}=r-R(\theta)
\end{eqnarray}
Hence we get the following expressions for the dissipation tensor $R_{ij}$:
\begin{eqnarray}
R_{i,j}=2^{|i-j|}\pi m\rho_0\overline v
\nonumber\\
\times\int_0^\pi\frac{\left\{R_0\left\{\frac{\partial\beta_1}{\partial\beta_i}Y_{10}+Y_{i0}\right\}+
\frac{\partial R_0}
{\partial\beta_i}S\right\}}{1+\frac{1}{R^2(\theta)}\left\{\frac{\partial R(\theta)}
{\partial\theta}\right\}^2}
\nonumber\\
\times \left\{R_0\left\{\frac{\partial\beta_1}{\partial\beta_j}Y_{10}+Y_{j0}\right\}+
\frac{\partial R_0}{\partial\beta_j}S\right\}  R^2(\theta)\sin\theta\,\mathrm{d}\theta
\end{eqnarray}

\subsubsection{The Langevin term}
The stochastic force $F_{fluc}(t)$    will 
determine fluctuations in momentum space, according to the value of the diffusion
coefficient $D$. We assume that
\begin{equation}
\langle F_{fluc}(t)F_{fluc}(t+s)\rangle=D\delta(s)
\end{equation}
The action of the stochastic force $F_{fluc}$ 
may be simulated numerically by repeatedly producing a random kick 
$\delta P$ in the
collective velocity associated with  the quadrupole and octupole moments.  
The value of $\delta P$ is chosen randomly from a Gaussian distribution 
with mean value and variance given by:
\begin{equation}
\overline{\delta P}=0
\end{equation}
\begin{equation}
\overline{(\delta P)^2}=D\delta t
\end{equation}
where $\delta t$ is the small time step between two kicks. 
The diffusion coefficient $D$ can be found using the Einstein relation:
\begin{equation}
D=2T\gamma,
\end{equation}
where $\gamma$ is the dissipation coefficient and $T$ is the 
temperature of the system \cite{Lan_book}. 
Hence the fluctuations that we are considering are induced essentially by the
thermal agitation. 
We notice that 
our dissipation tensor $R_{ij}$, introduced above,  has also nondiagonal 
terms. Hence, to correctly extract the dissipation coefficients,   
we diagonalaze the dissipation tensor 
$R_{ij}\to\gamma_{ij}$. The tensor $\gamma_{ij}$ will have 
only diagonal elements: $\gamma_2$ and $\gamma_3$. 
Now we can find $D_2$
and $D_3$ in the new coordinate system and evaluate 
$\delta P_2$ and $\delta P_3$, the random
kicks for the new coordinates. Finally it is possible to 
go back to the general coordinates $\beta_2$ and $\beta_3$, 
by the inverse transformation,  
and obtain $\delta P_{\beta_2}$ and $\delta P_{\beta_3}$.

\section{Results}

We will exploit the Langevin treatment outlined above to investigate the competition
between (incomplete) fusion and binary break-up mechanisms in low energy reactions. 
We consider the system $^{36}$Ar + $^{96}$Zr at two beam
energies, 9 and 16 MeV/u, in the following range of impact parameters: 
b = 5-7 fm and b = 4-6 fm at 9 and 16 MeV/u, respectively. 
Within this selection, according to the SMF dynamical evolution, 
one observes the formation of rather elongated configurations for which 
fluctuations are expected to be crucial in determining the following evolution. 
For lower impact parameters, the conditions of the
reactions are such that one always obtains incomplete fusion, while for larger impact
parameters bynary break-up is observed.
Contour plots of the density in the reaction plane, as obtained in the SMF
calculations, are displayed in Figs.1-2, for the two reactions. 

\begin{figure*}[t]
\includegraphics[width=9.cm]{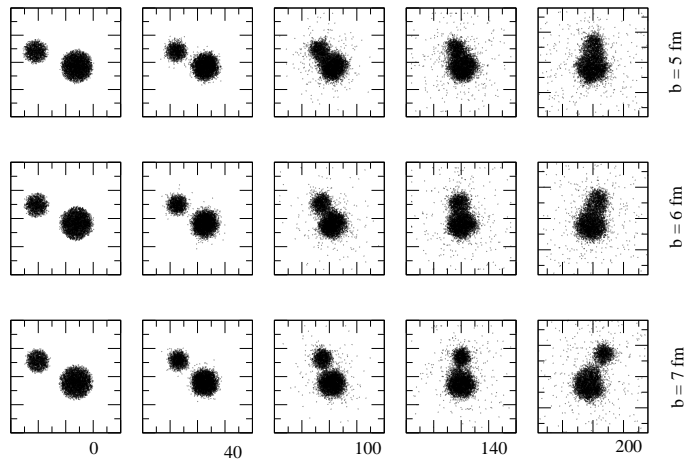}
\caption{Contour plots of the density projected on the reaction plane
  calculated with SMF for the reaction $^{36}$Ar +
  $^{96}$Zr at 9 MeV/u, at several times (fm/$c$). 
The size of each box is 40 fm.}
\label{contour_BGBD}
\end{figure*}
\begin{figure*}
\vskip 1.5cm
\includegraphics[width=9.cm]{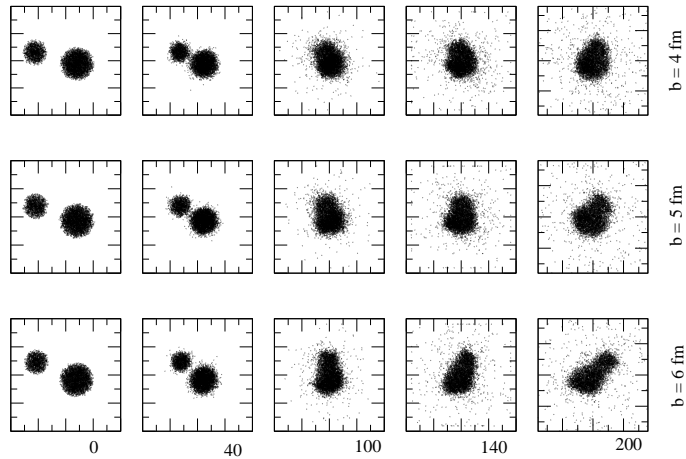}
\caption{The same as in Fig.\ \ref{contour_BGBD} but at 16 MeV/u.}
\label{contour_AMD}
\end{figure*}

The description of the system in terms of the global observables
$\beta_2$, $\beta_3$ and $\omega$
begins at the moment when, according to the full SMF evolution, the
composite system reaches a
quasi-stationary shape, having dissipated almost completely the radial
part of the kinetic
energy deposited into the system, while the angular part is converted into
intrinsic spin.
This time instant is estimated to be around $t_{freeze-out} \approx 200~fm/c$.
During the earlier dynamical evolution, pre-equilibrium nucleon emission takes
place. As a consequence, mass and charge of the system are smaller than
the total mass and charge numbers, respectively. We get $A\approx 122$, $Z\approx 53$. 
As one can see from Figs.1-2, the system configuration can be suitably parametrized
in terms of quadrupole and octupole moments.
From this point of view, the Langevin treatment introduced above appears
appropriate to describe the following evolution, 
though the dynamical description is devolved to few leading
degrees of freedom.
The initial conditions of the Langevin equation have been determined running 10 SMF trajectories.
The corresponding parameters are listed in
Tables I-II, for a couple of events, for each considered case.  

Then, within the Langevin treatment, 200 stochastic events 
were considered for 
each SMF trajectory. Fluctuations are injected each $3~fm/c$.  



\begin{table}[h]
\begin{tabular}{|c|c|c|c|c|c|c|c|}
\hline
b (fm) & $E^*$ (MeV) & L ($\hbar$) & $\beta_2$ & $\beta_3$ &
$\frac{d\beta_2}{dt}$ & $\frac{d\beta_3}{dt}$ & P \\
\hline
7 & 225 & 100 & 1.14 & -0.73 & 0.099 & 0.024 & 0.990 \\
\hline
7 & 242 & 95 & 1.00 & -0.76 & 0.143 & -0.129 & 0.990 \\
\hline
6 & 240 & 77 & 0.83 & 0.47 & 0.062 & -0.010 & 0.645 \\
\hline
6 & 224 & 84 & 1.01 & -0.52 & 0.113 & -0.063 & 0.880 \\
\hline
5 & 216 & 64 & 0.58 & -0.32 & 0.125 & 0.938 & 0.375 \\
\hline
5 & 227 & 58 & 0.56 & 0.36 & -0.004 & 0.005 & 0.145 \\
\hline
\end{tabular}
\caption{Characteristics of the composite system, as obtained in the reaction
 $^{36}$Ar + $^{96}$Zr at 9 MeV/u
at the time $t_{freeze-out}$: Excitation energy, intrinsic angular momentum, quadrupole
moment, octupole moment and associated collective velocities. 
The time 
unit adopted to define the  collective velocities is $10^{-22} s  = 30~ fm/c$.
Two events are
displayed for each impact parameter. The fission probability (see text) is reported
in the last column.}

\end{table}

\begin{table}[h]
\begin{tabular}{|c|c|c|c|c|c|c|c|}
\hline
b (fm) & $E^*$ (MeV) & L ($\hbar$) & $\beta_2$ & $\beta_3$ &
$\frac{d\beta_2}{dt}$ & $\frac{d\beta_3}{dt}$ & P \\
\hline
6 & 279 & 90 & 0.88 & 0.34 & 0.016 & 0.059 & 1.000 \\
\hline
6 & 277 & 97 & 0.88 & 0.44 & -0.015 & -0.031 & 1.000 \\
\hline
5 & 241 & 73 & 0.37 & 0.15 & -0.063 & -0.047 & 0.320 \\
\hline
5 & 252 & 77 & 0.63 & 0.40 & 0.136 & -0.020 & 0.580 \\
\hline
4 & 258 & 63 & 0.31 & 0.06 & 0.052 & 0.018 & 0.110 \\
\hline
4 & 247 & 52 & 0.22 & 0.05 & -0.007 & 0.002 & 0.035 \\
\hline
\end{tabular}
\caption{Same as in Table I, but for the reaction at 16 MeV/u.}
\end{table}

According to the values listed in Tables I-II, we test essentially the behavior
of composite systems with a variety of conditions of angular momentum, 
ranging from 50 $\hbar$ to 100 $\hbar$ and quadrupole moment $\beta_2$,
from 0.2 to 1. 
The excitation energy is about 250 MeV, corresponding to temperatures of the
order of 4 MeV. 
Apart from the situation observed in the case of b = 7 fm, E/A = 9 MeV/u,
the octupole moment, $\beta_3$, always takes rather small values, of both signs,  
indicating that the memory of the entrance channel mass asymmetry is lost. 
Also the quadrupole and octupole collective velocities are rather small and may
take values of both signs, suggesting that collective motions, apart from the
rotation associated with the intrinsic spin, are damped. 
These conditions correspond closely to quasi-stationary, metastable situations,
i.e. the system is stable against small shape fluctuations. From one side,  
it may evolve radiating its excitation energy and spin and relaxing slowly 
towards the spherical configuration. 
On the other hand, if the amplitude of the kicks of the associated collective velocities is
large enough, the system may overcome the fission barrier and 
reach configurations corresponding to
surface instabilities, from which it rapidly separates in two pieces.
However, one should also consider that 
the latter possibility is in competition with 
nucleon emission, that reduces the
excitation energy (and the associated amplitude of thermal fluctuations), 
while the shape of the system is evolving.   The nucleon emission rate can be evaluated according
to the standard Weisskopf formalism \cite{Sur}.
For the situations under study,
the excitation energy reduces, due to nucleon emission, 
approximately by 2.5 MeV each 30 fm/c. 
We follow the trajectory  of the system until the available excitation energy is fully
dissipated.

Hence, thanks to the introduction of fluctuations in the dynamical evolution, 
for a given impact parameter one observes a bifurcation of
trajectories, leading either to compact
shapes (fusion) or to elongated shapes, with large values of quadrupole
and/or octupole moments, that eventually cause the break-up of the system.
Actually the two possible outcomes are associated with a kind of bimodal
behavior of the shape observables, related to configurations corresponding 
to local minima of the total (surface + Coulomb) energy.
In may be interesting to notice that bimodality  has been recently
observed also in the context of liquid-gas phase transitions, where
volume instabilities are concerned and dilute systems may either
recompact to normal density or split into a huge number of small fragments \cite{Eric}. 
  
\subsection{Fission rates}
In the following, we will first discuss some illustrative results obtained in the
case of the reaction at 16 MeV/u, b = 5-6 fm. 
\begin{figure}
\includegraphics[width=20pc]{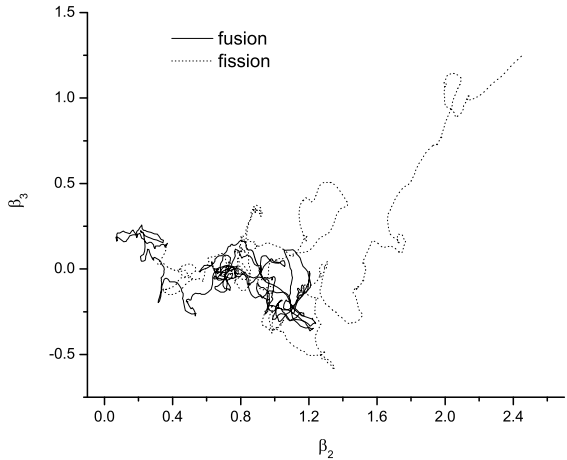}\hspace{2pc}%
\caption{\label{imb_eloss}
One example of trajectories leading either to fusion or
to break-up, in the ($\beta_2$,$\beta_3$) plane, as
obtained in the reaction at 16 MeV/u, b = 5 fm.} 
\end{figure}  
In Fig.3 we present one example of trajectories corresponding to the
two possibile exit channels (fusion or fission), in the ($\beta_2$, $\beta_3$) plane. 
Due to the random kicks, starting from the same initial conditions, rather
different paths are explored. 
It should be noticed that, also in the case of trajectories leading to
fusion, the final configuration is
not exactly spherical, but is associated with small (not vanishing) values of the quadrupole
moment.  
This corresponds to the stationary configuration
compatible with the amount of intrinsic angular momentum present in the system. 
On the other hand,  break-up configurations 
are characterized by rather large values of $\beta_2$ and/or $\beta_3$. 
Actually 
one sees an interesting 
correlation between the two parameters, that is
represented in Fig.4.   
In fact, both a
large quadrupole or octupole moments are linked to 
break-up configurations, that correspond to tangent spheroids. 
Fluctuations of the octupole moment are rather large, though the majority
of the events is located near $\beta_3 = 0$, corresponding to symmetric
fission.

\begin{figure}
\includegraphics[width=20pc]{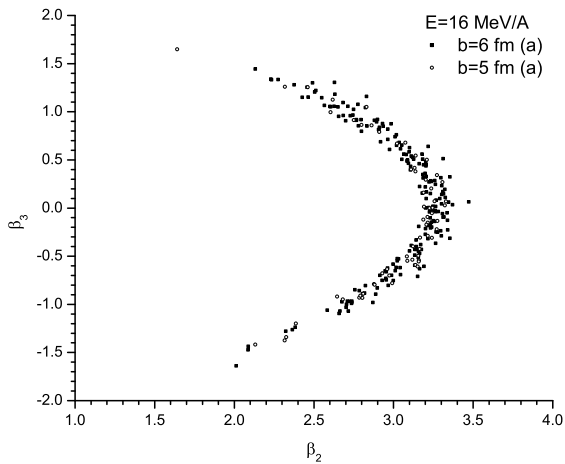}\hspace{2pc}%
\caption{\label{imb_eloss}
Correlations between the values of quadrupole and octupole moments, as obtained for
the break-up configurations in the case of the reaction at 16 MeV/u, b = 5-6 fm.} 
\end{figure}  

In Figs.5-6 (left) the fission rate, $dN/dt$, 
as obtained for
$b = 6-7~fm$ at $9~MeV/u$ and
$b = 5-6 fm$ at $16~MeV/u$, 
is displayed as a function of time for a set of 200 events in each of the 
cases considered.   
For the most peripheral impact parameters, 
after an initially increasing trend, related to the time
interval  needed to build and propagate fluctuations, we observe 
an almost
exponential decrease, as expected in the case of constant break-up probability $\gamma_{break}$. In  
this case one can write:
$dN/dt = N_t\gamma_{break}$, with $N_t = N_0 e^{-\gamma_{break} t}$ and $N_0 = 200$ (the total number of events
considered). 
This corresponds to situations where the break up probability ($\gamma_{break} \approx~ 0.002~ c/fm$) 
is not much affected by the competing nucleon emission. 
All events practically lead to fission 
over a time interval that is shorter than the one needed to exhaust the available
excitation energy by nucleon emission.
In fact, the maximum of the emission rate is observed at about 300 fm/c
and the system needs, on average, roughly 500 fm/c to reach the
break-up configuration (this is actually the half life time $\tau_{break} = 1/\gamma_{break}$). 
On the other hand, for smaller impact parameters (corresponding to lower deformation of the system
and lower angular momentum),
the break-up probability $\gamma_{break}$ is quenched approximately by a factor 4 
(see the left panel of Figs.5-6) and decreases in the course of time because of 
nucleon emission, that reduces 
the excitation energy and the corresponding amount of thermal fluctuations. 
In most cases, the excitation energy deposited into
the system is dissipated before the break up configuration may be reached. 
It is interesting to notice that, even in the most favourable case, the typical times of the process are rather 
large (500 fm/c), as compared for instance to the time scales associated with  the development
of volume instabilities in multifragmentation processes at higher energies (about 150 fm/c).  This 
can be explained in terms of the larger amount of excitation energy deposited into the system 
in the latter case (that induces fluctuations of higher amplitude and collective radial expansion)
and of the smaller growth times associated with volume instabilities \cite{rep1}.

The corresponding fraction of events that undergo break-up, $P_{break}$, is reported in Table I-II,
at the two energies and for all impact parameters considered.
From the estimated break-up probabilities it is possible to construct
the fusion cross-section, 
$\sigma_f(b) = (1-P_{break})~ 2\pi bdb$, that is displayed in Fig.7, for the two energies. 
We also show, for comparison, the results obtained within the SMF approach only, where 
due to the approximate treatment of fluctuations, one gets
distributions close to a sharp cut-off (approximated by a sharp cut-off in the figure). 
It is interesting to notice that, especially in the case of the reaction at 9 MeV/u,
the fusion cross section is reduced significantly by the introduction of fluctuations
that, in turn, help the system to overcome the fission barrier and to break-up.  

\begin{figure}[t]
\includegraphics[width=20pc]{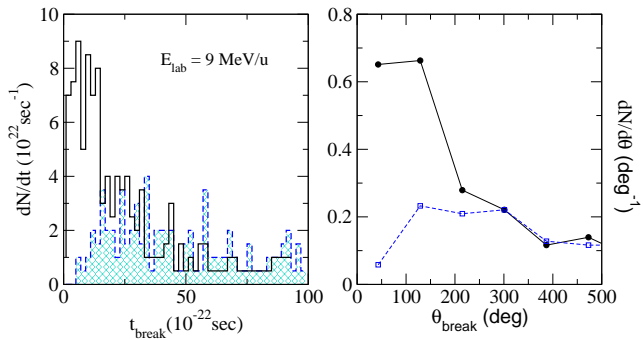}\hspace{2pc}%
\caption{\label{imb_eloss}
(Color online)
Left panel: Distribution of the time $t_{break}$ (see text), as obtained for the
reaction at 9 MeV/u and impact parameters b = 7 fm (full histogram)
and b = 6 fm (shaded histogram).
Right panel: Angular distribution of the break-up direction. Full line and circles refer
to b = 7 fm; dashed line and open squares are for b = 6 fm.}   
\end{figure}  

\begin{figure}[t]
\vskip 2.5cm
\includegraphics[width=20pc]{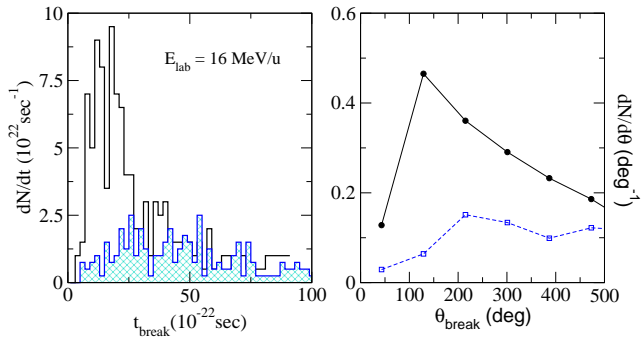}\hspace{2pc}%
\caption{\label{imb_eloss}
(Color online)
Same as Fig.5, for the reaction at 16 MeV/u and b = 5-6 fm.}
\end{figure}  

\subsection{Features of fission fragments}
The time $t_{break}$, needed to reach 
the break-up configuration, is connected to other interesting features of the reaction dynamics, 
depending on the various entrance channel conditions. In fact, 
due to the intrinsic spin, the system rotates while its shape evolves according to Eq.(4). As a consequence, 
the direction along which the system separates into pieces 
is strictly connected to $t_{break}$.
Hence the shape of the angular distribution of fission fragments can be used as a clock
of the collision, from which one can extract information 
on the break-up probability and the underlying reaction mechanism.  
This is an appealing issue that can be investigated also experimentally by looking at the angular
distribution of the emerging reaction products and at the possible existence 
of alignment effects \cite{Isa,Def}.
In the case of a fast break-up (fast fission) 
the angular distribution should exhibit a peak: 
due to the elongated shape of the system, the emission is not isotropic.
Along the separation
process, fragments acquire velocities essentially due to the Coulomb repulsion, according
to the Viola systematics, like in standard fission, but with a preferential emission axis.  
The distribution of the angle, $\theta_{break}$,
corresponding to the rotation (on the plane perpendicular to the direction of the intrinsic spin 
of the system) until the break-up configuration is reached 
is shown in Figs.5-6 (right panel), for the two energies and two impact parameters.
Obviously, the shape of this distribution depends on the fission probability, but also on the system angular
velocity (that in turn depends on the intrinsic spin). In fact, in absence of rotation (vanishing spin) 
the fragments would always be emitted along a fixed axis. 
In the case of the most peripheral events, a clear peak is observed in the distribution. 
On the other hand, for more central impact parameters, the half life time 
is much larger and one essentially gets a flat 
distribution for $\theta_{break}$, similarly to what is expected in the case of standard statistical fission.

\begin{figure}[b]
\vskip1.5cm
\includegraphics[width=18pc]{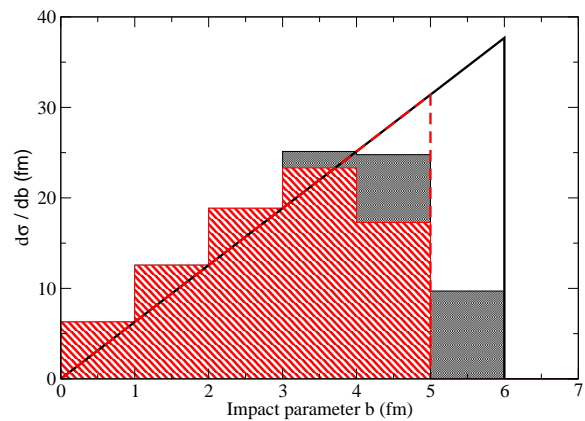}\hspace{2pc}%
\caption{\label{imb_eloss}
(Color online)
Fusion cross section, as a function of the impact parameter b, as obtained in the reactions
at 9 MeV/u (black histogram) and 16 MeV/u (grey histogram) with the Langevin treatment, Eq.(4).
The lines correspond to SMF simulations at 9 MeV/u (full) and 16 MeV/u (dashed).}
\end{figure}  
 \begin{figure*}[t]
\includegraphics[width=12.cm]{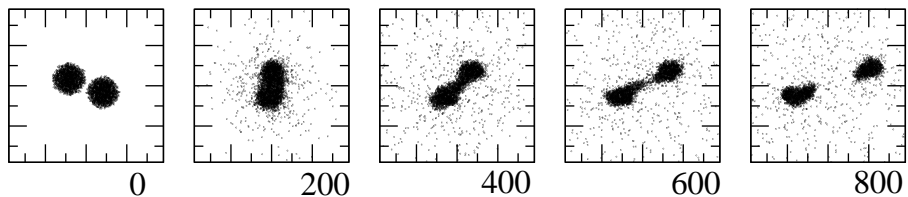}
\caption{Contour plots of the density projected on the reaction plane
  calculated with SMF for the reaction $^{197}$Au +
  $^{197}$Au at 15 MeV/u, b = 6 fm, at several times (fm/$c$). 
The size of each box is 80 fm.}
\label{contour_AMD_1}
\vskip2.0cm
\end{figure*}

\subsection{Fast-fission of PLF and TLF}
Several shape, angular momentum and excitation energy conditions  
can be observed also, in the case of collisions between heavy systems, after the separation
into PLF and TLF, for one (or both) of these products. Thus it is interesting to investigate fast fission processes
of these objects, leading to ternary (or quaternary) breaking of the whole system.  
For instance, we display in Fig.8, density contour plots as 
obtained in SMF simulations of semi-peripheral collisions of
Au + Au at 15 MeV/u, for which aligned ternary and quaternary breaking has been recently observed 
experimentally \cite{Isa}. 
One can see that similar shape configurations, as the ones observed in the
reactions investigated above, may appear for PLF/TLF fragments.
However, these fragments have lower angular momentum (about $20-40\hbar$) and excitation energy (of the order of 100 MeV).
The corresponding break-up probability is of the order of $10\%$ and emission times are longer
($\approx~$ 2000 fm/c).  
The fast-fission mechanism could explain qualitatively some of the features observed experimentally, such as
alignment effects and fragment relative velocities and charge distributions.
However, a thourough analysis of the kinematical properties of the reaction products \cite{Isa}, 
as well as the estimated rather short break-up times,
suggests 
the persistence of non-equilibrium effects in momentum space, i.e. the presence of collective  
velocities in $\beta_2$ and/or $\beta_3$, 
in addition to the tangential velocity generated by the intrinsic angular momentum.
Collective velocities, probably underestimated in the SMF calculations, would speed
up the fragmentation process since the system is pushed towards more exotic shapes, from which
it is easier to overcome the fission barrier.


\section{Conclusions}
In this article we have investigated the role of shape fluctuations in the
dynamical evolution of excited systems that can be formed in semi-peripheral
reactions at low energy (around 20 MeV/u). Quasi-stationary composite systems, with 
quadrupole and/or octupole deformation, are observed, for which shape fluctuations 
are essential to overcome the fission barrier and  eventually break-up.
This analysis is performed within a hybrid treatment that couples the
study of the early stage of the dynamics, devolved to a microscopic stochastic transport approach,
up to the formation of primary excited sources, 
to a full Langevin description of the leading degrees of freedom of these objects: quadrupole,
octupole moments and angular velocity. 
For temperature, shape and angular momentum conditions obtained in semi-peripheral reactions,
typical time scales of the break-up process are of the order of 500 fm/c.
The fission fragments are emitted along a preferential direction, that corresponds
to the maximum elongation axis. Due to angular momentum effects, this direction
may rotate while the shape of the system is evolving towards break-up configurations.
Hence a careful analysis
of the angular distribution of the reaction products may give
relavant information about fission probabilities and the involved
time scales, that in turn are closely linked to the mean-field
dynamics and the properties of the nuclear interaction (range, surface energy, 
two-body correlations). 
From this study it is clear that a good treatment of mean-field fluctuations
is a crucial point in the characterization of dissipative reactions. 
The model employed here provides a suitable 
description of surface modes, parametrized in terms of quadrupole and
octupole  oscillations, but it could miss some non-equilibrium effects 
that can help the system to break-up.
In fact, 
collective velocitities related to shape observables 
are likely underestimated in the SMF approach 
\cite{frag}
and the role of multipolarities higher than octupole is neglected in the
Langevin treatment. 
A fully microscopic description of the whole process would be highly
desirable, though it is far from being trivial.
Some attempts are represented by improuved quantum
molecular dynamics calculations (ImQMD) \cite{QMD}. 
Stochastic extensions of Time-Dependent-Hartree-Fock (TDHF) calculations
should also provide a valuable tool to characterize reaction
mechanisms in  low energy collisions  \cite{Denis}. 
Work is in progress in this direction.



\end{document}